\documentclass{webofc}
\usepackage[varg]{txfonts}   % Web of Conferences font
\usepackage{verbatim}
\begin{document}
\title{Improved quark coalescence model for spin alignments of vector mesons}
%
% subtitle is optionnal
%
%%%\subtitle{Do you have a subtitle?\\ If so, write it here}

\author{\firstname{Xin-Li} \lastname{Sheng}\inst{1}\fnsep\thanks{\email{xls@mail.ustc.edu.cn}}
        % etc.
}

\institute{Key Laboratory of Quark and Lepton Physics (MOE) and Institute of Particle Physics, Central China Normal University, Wuhan, Hubei 430079, China.
          }

\abstract{%
We propose an improved quark coalescence model for spin alignment of vector mesons by spin density matrix with phase space dependence. Within this model we propose an understanding of spin alignments of vector mesons $\phi$ and $K^{*0}$ in the static limit: a large positive deviation of $\rho_{00}$ for $\phi$ mesons from $1/3$ may come from the electric part of the vector $\phi$ field, while a negative deviation of $\rho_{00}$ for $K^{*0}$ mesons may come from the electric part of vorticity fields. In the low-$p_T$ region, $\rho_{00}$ for $K^{*0}$ mesons is proportional to $p_T^2$, which is qualitatively agree with experimental results.
}
\maketitle
\section{Introduction}
\label{intro}
In ultra-relativistic heavy-ion collisions, a huge orbital angular momentum (OAM) can be generated in the direction perpendicular to the reaction plane. During the evolution of the fireball, the OAM is transferred to the spin polarization of quarks through the spin-orbit coupling in nonlocal scatterings \cite{Liang:2004ph}. The polarized quarks will then combine and form polarized hadrons and mesons in the freezeout stage \cite{Liang:2004ph,Liang:2004xn,Yang:2017sdk}. The spin alignment of vector mesons can be measured through the angular distribution of decay daughters \cite{Schilling:1969um}. Here we report on our recent paper \cite{Sheng:2020ghv}, where an improved quark coalescence model in phase space with spin degrees of freedom was proposed. We give a general relation between the density matrix of vector mesons and that of quarks. Using the spin polarization of quarks in vorticity and vector meson fields, predicted by the Wigner function approach \cite{Yang:2017sdk}, we derive spin alignments of vector mesons $\phi$ and $K^{*0}$. Our work propose possible explanations for the experimental results obtained by ALICE and STAR collaborations \cite{ALICE:2019aid,Kundu:2021lra,Singha:2020qns}. This work should be tested by a detailed and comprehensive simulation
of vorticity tensor fields and vector meson fields in heavy ion collisions.

{\it Notations and conventions. } We suppress the time dependence of all quantities throughout this paper since we work at the formation time of hadrons. We use boldfaced symbols to represent three-vectors, e.g., ${\bf x}$ or ${\bf p}$, and use ${\bf x}_i$ with subscript $i=x,y,z$ for the three components. The momentum integration measure is denoted using shorthand notation $[d^3{\bf p}]\equiv d^3{\bf p}/(2\pi)^3$.

\section{Spin density matrix and quark coalescence model in phase space}
\label{sec-1}
In Ref. \cite{Yang:2017sdk}, a quark coalescence model is constructed based on the spin density matrix in momentum space. However,
we need to formulate a model that includes space-time dependence because spin polarizations of quarks are inhomogeneous in heavy ion collisions. We generalize the definition of density operator by including off-diagonal elements in momentum space,
\begin{equation}
\rho = \sum_s \int d^3 {\bf x} \int [d^3{\bf p}]w(s,{\bf x},{\bf p})\int[d^3{\bf q}]e^{-i{\bf x}\cdot{\bf q}}
\left|s,{\bf p}+\frac{{\bf q}}{2}\right\rangle\left\langle s, {\bf p}-\frac{{\bf q}}{2}\right| \,, \label{density operator}
\end{equation}
where $\left|s,{\bf p}\right\rangle$ is the spin-momentum state of considered particles. Here ${\bf x}$ is the conjugate position of ${\bf q}$, i.e., the difference between momenta of state bases. Such a definition is consistent with the wave-packet description for a quantum particle, wherein a particle has finite momentum uncertainty and consists of plane waves with various momenta.
The weight function $w(s,{\bf x},{\bf p})$ is actually the Wigner function, which is the probability of finding a particle (described by a wave packet) with spin $s$, center position ${\bf x}$, and average momentum ${\bf p}$. It can be obtained by projecting the density operator onto two states with the same spin and different momenta
\begin{equation}
w(s,{\bf x},{\bf p})=\int[d^3{\bf q}]e^{i{\bf q}\cdot{\bf x}}\left\langle s,{\bf p}+\frac{{\bf q}}{2}\right|\rho\left|s,{\bf p}-\frac{{\bf q}}{2}\right\rangle \,. \label{weight function}
\end{equation}
For quarks with spin $1/2$, we parameterize the weight function as
\begin{equation}
w(q|s,{\bf x},{\bf p})=\frac{1}{2}f_q({\bf x},{\bf p})\left[1+sP_q({\bf x},{\bf p})\right] \,,\label{fermion weight function}
\end{equation}
where $f_q({\bf x},{\bf p})$ is the normalized quark distribution function, $\int d^3{\bf x}\int d^3[{\bf p}] f_q({\bf x},{\bf p})=1$, and $P_q({\bf x},{\bf p})$ denotes the average spin polarization at phase space point $({\bf x},{\bf p})$.

We note that definitions of the density operator in Eq. (\ref{density operator}) and the weight function in Eq. (\ref{weight function}) can be applied to any kind of particles, such as quarks/antiquarks or vector mesons. Therefore spin density matrix elements for mesons are obtained similar to Eq. (\ref{weight function}) by putting the density matrix operator between two meson states,
\begin{equation}
\rho^M_{S_{z1},S_{z2}}({\bf x},{\bf p})=\int [d^3{\bf q}]e^{i{\bf q}\cdot{\bf x}}\left\langle M;S,S_{z1};{\bf p}+\frac{\bf q}{2}\right| \rho_M \left|M;S,S_{z2};{\bf p}-\frac{\bf q}{2}\right\rangle \,, \label{rho meson}
\end{equation}
where $M$ labels the flavor state of the meson, $S=1$ for a vector meson, and $S_z=-1,0,1$ denote spin states in the spin quantization direction. We take the approximation that the meson's density matrix is given by the direct product of density matrices of its constitute quark/antiquark, $\rho_M=\rho_{q\bar{q}}\equiv \rho_q \otimes\rho_{\bar{q}}$. Putting $\rho_{q\bar{q}}$ into Eq. (\ref{rho meson}), we derive
\begin{align}
\rho^M_{S_{z1},S_{z2}}({\bf x},{\bf p})=& \int d^3{\bf x}_b [d^3 {\bf p}_b][d^3 {\bf q}_b] e^{-i{\bf q}_b \cdot {\bf x}_b} \varphi_M^* \left({\bf p}_b+\frac{{\bf q}_b}{2}\right)\varphi_M \left({\bf p}_b-\frac{{\bf q}_b}{2}\right) \nonumber \\
& \times \sum_{s_1,s_2} w\left( q_1\big | s_1,{\bf x}+\frac{{\bf x}_b}{2},\frac{{\bf p}}{2}+{\bf p}_b\right) w\left(\bar{q}_2\big | s_2,{\bf x}-\frac{{\bf x}_b}{2},\frac{{\bf p}}{2}-{\bf p}_b\right) \nonumber \\
& \times \left\langle S,S_{z1}\,\big|\, s_1,s_2\right\rangle \left\langle s_1,s_2\,\big|\,S,S_{z2}\right\rangle \,, \label{meson rho}
\end{align}
where we have used the fact that the spin state in a meson's wave function can be decoupled from the momentum state. The combination of spin states is controlled by the Clebsch-Gordan coefficients
$\langle S,S_{z1}\,|\, s_1,s_2\rangle$. Meanwhile, the inner product between the meson's momentum state $\left|M;{\bf p}\right\rangle$ and the quark-antiquark's momentum state $\left|{\bf p}_1,{\bf p}_2\right\rangle$ gives the meson wave function $\varphi_M$,
\begin{equation}
\left\langle{\bf p}_1,{\bf p}_2\big|M; {\bf p}\right\rangle = (2\pi)^3 \delta^{(3)}\left({\bf p}_1+{\bf p}_2-{\bf p}\right)\varphi_M\left(\frac{{\bf p}_1 - {\bf p}_2}{2}\right) \,.
\end{equation}
The simplest choice of $\varphi_M({\bf k})$ is the Gaussian distribution $\varphi_M({\bf k})=N_M \exp[-{\bf k}^2/(2a_M^2)]$, where $a_M$ is the average momentum width for the meson's wave function. The pre-factor $N_M$ is determined by the normalization condition $\int[d^3{\bf k}]\varphi^*_M({\bf k})\varphi_M({\bf k})=1$.

\section{Spin alignments of $\phi$ and $K^{\ast 0}$}
\label{sec-2}
In relativistic heavy ion collisions, there are different sources for spin polarizations of massive fermions: vorticity fields, electromagnetic fields, and mean fields of vector mesons. Here we only consider polarizations induced by vorticity and vector meson fields since contributions from electromagnetic fields are believed to be negligible \cite{Sheng:2019kmk}. For quarks and antiquarks, the polarization along the direction of global OAM, i.e., the $y$-direction, are given by
\begin{equation}
P^y_{q/\bar{q}}({\bf x},{\bf p})=\frac{1}{2}\boldsymbol\omega_y \pm \frac{1}{2m_q}(\boldsymbol{\varepsilon}\times{\bf p})_y\pm\frac{g_V}{2m_q T}{\bf B}_y^V+\frac{g_V}{2m_q E_p T}({\bf E}_V\times{\bf p})_y \,, \label{quark polarization}
\end{equation}
which are derived in the Wigner function approach \cite{Yang:2017sdk}. We have defined the three-vector of the electric part of the thermal vorticity tensor as $\boldsymbol\omega = \nabla\times(\beta{\bf u})/2$, and the magnetic part as $\boldsymbol\varepsilon=-[\partial_t(\beta{\bf u})+\nabla(\beta u^0)]/2$, where $\beta=1/T$ is the inverse temperature and $u^\mu=(u^0,{\bf u})$ is the flow velocity. The strength tensor $F_V^{\mu\nu}$ of vector meson fields is also decomposed into the electric part ${\bf E}^V_i=F_V^{i0}$ and the magnetic part ${\bf B}^V_i=-\epsilon^{ijk}F_{V}^{jk}$ in analogy with normal electromagnetic fields, with $i,j,k=x,y,z$. We use $g_V$ to represent the coupling constant between quark/antiquark and vector meson fields in the quark-meson model \cite{Sakurai:1960ju,Zacchi:2015lwa}. For light quarks, i.e., $u$ and $d$ quarks, the major contributions are $\rho$ and $\omega$ meson fields, while $s$ quarks are mainly polarized by $\phi$ meson fields.

Substituting the average polarization (\ref{quark polarization}) for $s$ and $\bar{s}$ into Eqs. (\ref{fermion weight function}) and (\ref{meson rho}), we obtain the average value of the normalized spin density matrix element $\bar\rho_{00}\equiv\rho_{00}/(\rho_{-1,-1}+\rho_{00}+\rho_{11})$ (which is also called the spin alignment) for $\phi$ mesons,
\begin{align}
\left\langle\bar\rho^{\phi}_{00}({\bf x},{\bf p})\right\rangle \approx & \frac{1}{3}-\frac{1}{9}\left\langle\boldsymbol\omega_y^2\right\rangle -\frac{1}{27m_s^2}\left(\left\langle\boldsymbol\varepsilon_x^2\right\rangle +\left\langle\boldsymbol\varepsilon_z^2\right\rangle\right) \left\langle{\bf p}_b^2\right\rangle_\phi \nonumber \\
& +\frac{g_\phi^2}{9m_s^2}\left[\left\langle\left(\beta {\bf B}_y^\phi\right)^2\right\rangle + \left\langle\left(\beta {\bf E}_z^\phi\right)^2\right\rangle \left\langle\frac{{\bf p}_{b,x}^2}{E_{p_1}E_{p_2}}\right\rangle_\phi+ \left\langle\left(\beta {\bf E}_x^\phi\right)^2\right\rangle \left\langle\frac{{\bf p}_{b,z}^2}{E_{p_1}E_{p_2}}\right\rangle_\phi \right] \nonumber \\
& +\frac{1}{36m_s^2}\left[\left\langle\boldsymbol\varepsilon_z^2\right\rangle -g_\phi^2\left\langle\left(\beta {\bf E}_z^\phi\right)^2\right\rangle \left\langle\frac{1}{E_{p_1}E_{p_2}}\right\rangle_\phi\right] {\bf p}_x^2 \nonumber \\
& +\frac{1}{36m_s^2}\left[\left\langle\boldsymbol\varepsilon_x^2\right\rangle -g_\phi^2\left\langle\left(\beta {\bf E}_x^\phi\right)^2\right\rangle \left\langle\frac{1}{E_{p_1}E_{p_2}}\right\rangle_\phi\right] {\bf p}_z^2 \,, \label{phi spin alignment}
\end{align}
where ${\bf p}_1={\bf p}/2+{\bf p}_b$ and ${\bf p}_2={\bf p}/2-{\bf p}_b$. Here the spin quantization direction is chosen as the $y$-direction. We use $\langle\cdots\rangle$ to represent the average over generating positions of $\phi$ mesons, while $\langle\cdots\rangle_\phi$ to represent the average on the $\phi$ meson's wave function. We have $\langle\cdots\rangle_\phi\equiv8\int d^3{\bf x}_b d^3 {\bf p}_b \exp\left(-{\bf p}_b^2/a_M^2-a_M^2 {\bf x}_b^2\right)(\cdots)$ if we take the Gaussian distribution approximation for the meson wave function. We find that, in Eq. (\ref{phi spin alignment}), all contributions appear independently as positive or negative quantities.
For nearly static $\phi$ mesons with ${\bf p}\simeq 0$, vorticity fields, including $\boldsymbol\omega$ and $\boldsymbol\varepsilon$, have negative contributions, while vector meson fields ${\bf E}_\phi$ and ${\bf B}_\phi$ have positive contributions. We have argued in Ref. \cite{Sheng:2019kmk} that the dominant contribution to $\bar{\rho}^\phi_{00}$ may possibly be from the electric part ${\bf E}_\phi$, which results in the positive deviation from $1/3$ for $\phi$ mesons' spin alignment \cite{ALICE:2019aid,Kundu:2021lra,Singha:2020qns}.

Let us turn to the spin alignment of vector meson $K^{*0}$ with flavor $(d\bar s)$. We note that $d$ quarks will be polarized by $\rho$ and $\omega$ meson fields, i.e., $P^y_d$ depends on ${\bf E}_\rho$, ${\bf B}_\rho$, ${\bf E}_\omega$, and ${\bf B}_\omega$. On the other hand, $\bar{s}$ is polarized by $\phi$ meson fields ${\bf E}_\phi$ and ${\bf B}_\phi$. We take the approximation that different kinds of meson fields do not have large correlation in space, therefore only quadratic terms survive in the spin alignment of $K^{*0}$,
\begin{align}
\left\langle\bar\rho^{K^{*0}}_{00}({\bf x},{\bf p})\right\rangle \approx & \frac{1}{3}-\frac{1}{9}\left\langle\boldsymbol\omega_y^2\right\rangle -\frac{1}{27m_s^2}\left(\left\langle\boldsymbol\varepsilon_x^2\right\rangle +\left\langle\boldsymbol\varepsilon_z^2\right\rangle\right) \left\langle{\bf p}_b^2\right\rangle_{K^{*0}} \nonumber \\
& +\frac{1}{36m_s m_d}\left[\left\langle\boldsymbol\varepsilon_z^2\right\rangle {\bf p}_x^2+\left\langle\boldsymbol\varepsilon_x^2\right\rangle {\bf p}_z^2 \right] \,.\label{K spin alignment}
\end{align}
This result is obtained by substitute polarizations of $d$ and $\bar{s}$, given by Eq. (\ref{quark polarization}), into Eqs. (\ref{fermion weight function}) and (\ref{meson rho}). We see that contributions from vector meson fields are absent in Eq. (\ref{K spin alignment}). The spin alignment of $K^{*0}$ is dominated by vorticity fields and will be smaller than $1/3$ for nearly static $K^{*0}$, which qualitatively agrees with experimental results obtained by ALICE and STAR \cite{ALICE:2019aid,Kundu:2021lra,Singha:2020qns}. We note that the contribution from $\boldsymbol\varepsilon$ for $\bar{\rho}_{00}^{K^{*0}}$ is amplified by about $2.1\sim2.3$ compared to that for $\bar{\rho}_{00}^{\phi}$. Therefore $\boldsymbol\varepsilon$ may have a sizable magnitude of contribution for $\bar{\rho}_{00}^{K^{*0}}$, but is less important for $\bar{\rho}_{00}^{\phi}$. We also notice that the spin alignment of $K^{*0}$ is proportional to $p_T^2$ in the low-$p_T$ region, which seems to be agree with the $p_T$ dependence in Refs. \cite{ALICE:2019aid,Kundu:2021lra,Singha:2020qns}.

\section{Summary}
We have constructed an improved quark coalescence model based on the spin density matrix in phase space with coordinate dependence. Spin alignments for vector mesons $\phi$ and $K^{*0}$ are derived from spin polarizations for quarks. We propose understandings of significant positive (negative) deviations of $\bar\rho_{00}$ for $\phi$ ($K^{*0}$) mesons from $1/3$, which should be tested by detailed simulations of vorticity and vector meson fields in heavy ion collisions.

\section*{acknowledgments}
X.-L. S. is supported in part by the National Natural Science Foundation of China (NSFC) under Grant Nos. 11935007, 11221504, 11861131009, 11890714 (a sub-grant of 11890710), 12047528, and by the Fundamental Research Funds for Central Universities in China by the UCB-CCNU Collaboration Grant.

%
% BibTeX or Biber users please use (the style is already called in the class, ensure that the "woc.bst" style is in your local directory)
% \bibliography{name or your bibliography database}
%
% Non-BibTeX users please use
%

\end{document}